\documentclass[12pt]{article}

\usepackage{amsmath}
\usepackage{xcolor}
\usepackage{framed}
\definecolor{shadecolor}{rgb}{0.90,0.90,0.90}
\usepackage{hyperref}
\usepackage{subcaption}
\usepackage[compat=1.1.0]{tikz-feynman}
\usepackage[utf8]{inputenc}

\usepackage{mathtools}
\usepackage{setspace}
\usepackage{amsmath, amssymb, amsthm, float, graphicx}
\numberwithin{equation}{section}

\hypersetup{colorlinks=false, linkcolor=blue, citecolor=red}

\def\beq{\begin{eqnarray}}\def\eeq{\end{eqnarray}}
\def\be{\begin{equation}}\def\ee{\end{equation}}
\def\nn{\nonumber}

\def\bz{\bar{z}}

\newcommand{\CV}{{\cal V}}

\begin{document}
\title{\bf  Virasoro blocks at large exchange dimension}
\date{}

\author{Carlos $\text{Cardona}^{1/h}$,\\${}^{1/h}$Department of Physics, Arizona State University, Tempe, Arizona 85287, USA
}

\maketitle
\vskip 2cm
\abstract{In this paper, we analyze Virasoro conformal blocks in the limit when the operator exchange dimension is taking to be large in comparison with the other parameters dependence of the block. We do this by using Zamolodchikov's recursion relations. We found a dramatically simplified solution at leading order in an inverse power expansion in large exchange conformal dimension in terms of a quasi-modular form in an Eisenstein series representation. We compare this solution with existing results obtained previously by using AGT correspondence.}

\tableofcontents
\vfill {\footnotesize 
	ccardon6@asu.edu, cargicar@gmail.com}
\onehalfspacing

\newpage

\section{Introduction}
Conformal field theories in two dimensions play a major role in fundamental physics  and are ubiquitous in modern theoretical high energy physics. They find applications across a wide range of topics and space-time dimensions from realistic phenomena to purely mathematically motivated models. Some examples worth mentioning are: condense matter systems at criticality, such as the Ising model and models within its universality class,  fractional quantum Hall effect, deep inelastic scattering, string theory as well as AdS/CFT correspondence. Within them we find the most successful application so far of the nowadays mainstream bootstrap program, in particular because this approach lead to the discovery of an important large family of solvable theories known as minimal models \cite{Belavin:1984vu}. They have become important in the study of partition functions in supersymmetric gauge theories through the so-called AGT correspondence \cite{Alday:2009aq, Fateev:2009aw} and more recently, appear to be related to the asymptotic symmetries of quantum gravity in four dimensions \cite{Barnich:2009se, Kapec:2014opa, Cardona:2017keg}. They have served multiple times as an inspiration for developments in pure mathematics and find applications, in for example, the theory of Riemann surfaces. 

The main observables in a conformally invariant theory are correlation functions, which admit an operator product expansion decomposition in terms of conformal blocks. In the two dimensional case, the symmetry algebra is infinitely large and is known as the Virasoro algebra. The representations of this algebra can in turn be decomposed as a direct sum of the so-called Verma modules, each of which contains the whole family of descendants states created from a given highest weight state. The contribution to a correlation function from an entire Verma module is contained in a corresponding Virasoro block \cite{Belavin:1984vu} and therefore they are completely determined by the conformal symmetry. By knowing them, we can therefore methodically isolate the symmetry constrains on a generic correlation function and we are then left with the computation of the expansion coefficients, or OPE coefficients, in order to completely solve a particular theory. Despite the importance of the Virasoro blocks for the study and use of conformal field theories, a close, generic, complete and useful expression for them has elude us so far \footnote{Remarkably, crossing symmetry has been solved with not need of an explicit representation of the Virasoro blocks \cite{Ponsot:1999uf, Ponsot:2000mt} }. 

Some few Virasoro blocks has been computed exactly for particular theories and correlation functions. For example in minimal models, conformal blocks for four-point functions simplify drastically into Gauss hypergeometric functions when the correlator includes at least one degenerate field at level two \cite{Belavin:1984vu}. For a particular correlator between four operators of the same weight and particular central charge $c =1$ and $c=25$, an exact expression is known from \cite{Zamolodchikov:426555} \footnote{Another nice example of a theory with $c=1$ has been discussed in \cite{Runkel:2001ng}}. Some few particular cases were obtained by relating Virasoro blocks to Painlev\'e VI equation \cite{Gamayun:2012ma, Iorgov:2013uoa, Iorgov:2014vla, Litvinov:2013sxa}.  By using the correspondence between correlators in Liouville theory and instanton partition functions in supersymmetric field theories conjectured in \cite{Alday:2009aq}, some combinatorial formulas for the coefficients in a series expansion on cross ratios have also been obtained \cite{Alba:2010qc} (see also  \cite{Perlmutter:2015iya} for some combinatorial formulas found from CFT principles alone). A particular vacuum block in a semiclassical limit (large$-c$)  has been computed, and extensively studied in a series of papers \cite{Fitzpatrick:2015foa,Fitzpatrick:2015qma, Fitzpatrick:2015zha, Fitzpatrick:2014vua, Fitzpatrick:2016mtp, Hijano:2015qja, Fitzpatrick:2015dlt, Beccaria:2015shq, Besken:2019jyw}.
 
One could compute Virasoro blocks from first principles by using OPE expansions and the Virasoro algebra, computing level by level the contributions to the Virasoro blocks as is usually taught in textbooks \cite{Ginsparg:1988ui, di1996conformal}, but this becomes cumbersome very quickly due to the highly combinatorial nature of this approach, so it would be helpful to have alternative ways to treat them. In a couple of beautiful papers, Zamolodchikov developed an alternative approach by deriving a couple of recursion relations for Virasoro blocks,  by viewing them as an expansion in poles in the central charge \cite{Zamolodchikov:1987}, and as an expansion in poles in exchange dimension \cite{Zamolodchikov:1985ie}. This recursions have been later generalized to other cases, like to torus one-point functions \cite{Hadasz:2009db,Poghossian:2009mk,Fateev:2009me, Alkalaev:2016fok}, and the so-called heavy-light semiclassical limit \cite{Fitzpatrick:2014vua}. Among the applications of this recursion relations we can mention a few where it has been used in determining combinatorial expression for the coefficients in  cross-ratio expansions \cite{Perlmutter:2015iya} and in performing numerics in the context of AdS/CFT correspondence in \cite{Chen:2017yze}

Motivated by the simplified expression for the Virasoro blocks from \cite{Fitzpatrick:2015zha, Fitzpatrick:2014vua, Fitzpatrick:2015dlt, Beccaria:2015shq} obtained by a perturbative semiclassical analysis in $c^{-1}$, we expect similar simplifications in a perturbative treatment in a large exchange conformal dimension $1/h$. It is the goal of this paper to provide such an example by solving (one of) Zamolodchikov's recursion relation at leading order in a $1/h$ expansion for large$-h$. We find that the leading order correction, is given in terms of a Eisenstein series quasi-modular form, which is in agreement with expected results from the study of Virasoro blocks over elliptic curves. Even more, this result had appeared before from the use of the so-called  AGT correspondence \cite{Alday:2009aq} that relates Virasoro blocks to instanton partition functions in supersymmetric gauge field theories. In particular, we noticed that the solution found in this paper is contained in the analysis of \cite{Kashani-Poor:2013oza, Billo:2013fi} by using AGT correspondence, and extensions of this method to the torus one-point function can be  
found at \cite{KashaniPoor:2012wb,Kashani-Poor:2014mua,Huang:2011qx,Billo:2011pr}.

This limit is important because, among other particularities, it controls a universal sector of all non-rational two dimensional conformal field theories, which in turn behaves Liouville-like as has been recently proved in \cite{Collier:2019weq}. This can be checked in a traditional bootstrap computation \cite{Cardona:2020-2} and some aspects of it has been considered from the point of view of the modular bootstrap in \cite{Brehm:2019pcx,Das:2017cnv}. In particular, for conformal field theories with holographic description, the Lioville-like universal behavior of two dimensional conformal field theories in the large exchange region has been argued some little ago \cite{Jackson:2014nla}. Therefore, we believe our results in this paper will find applications in the study of this universal sector.

\section{Virasoro conformal blocks}
The spectrum of states for two dimensional conformal field theories fall into representations of the Virasoro algebra generated by the Laurent modes of infinitesimal conformal transformations. Due to the fact that in two dimensions, local conformal transformations corresponds to arbitrary analytic functions, the generic Laurent expansion of such functions will be in principle expanded by an infinite number of modes, and henceforth the number of generators in two dimensions is infinitely large. The Virasoro algebra is explicitly given,
\be
[L_m,L_n] = (m-n)L_{m+n} + {c\over 12}n(n^2-1)\delta_{m+n,0}~.
\ee
for the holomorphic sector and a similar algebra for the anti-holomorphic one. Modes with negative integer label corresponds to raising operators whereas positive label modes corresponds to lowering operators. The spectrum is build upon acting with raising operators over eigenstates of the zero mode, corresponding to  local primary operator by means of the state-operator correspondence, defined by,
\be
L_0{\cal O}_{h} |0\rangle\equiv L_0|{\cal O}_{h} \rangle=h |{\cal O}_{h} \rangle\,.
\ee
States created out of the primary states by the application of raising operators are known as descendants and the set of descendants from a given primary is known as a Verma module. The spectrum is then classified as a direct sum of all Verma modules.

Let us now consider a four-point correlation function of primary operators $\langle \prod_{i=1}^4{\cal O}_{h_i}(z_i)\rangle$. Conformal invariance greatly constrains the form of this observable, as being proportional to a function of the cross ratios only,
\be
z={z_{12}z_{34}\over z_{13}z_{24}},\quad \bz={\bz_{12}\bz_{34}\over \bz_{13}\bz_{24}}\,,
\ee
where we have used $z_{ij}\equiv z_i-z_j$. Explicitly, it is constrained to be of the form,
\be 
\langle \prod_{i=1}^4{\cal O}_{h_i}(z_i)\rangle=G(z,\bar{z})\prod_{i<j}^4z_{ij}^{{\sum h_i\over 3}-h_i-h_j}\bar{z}_{ij}^{{\sum \bar{h}_i\over 3}-\bar{h}_i-\bar{h}_j}\,.
\ee
we can additionally use global conformal invariance to fix three of the primary positions, namely $z_1=\infty, z_2=1, z_4=0$, which lead us to $z_3=z$, and similarly for the bared coordinates.

By taking the limit $(z,\bz)\to (0,0)$, the function $G(z,\bar{z})$ can be expanded as,
\be 
G(z,\bz)=\sum_{h,\bar{h}} C_{12,h, \bar{h}}C^{h, \bar{h}}_{~34} |{\cal }F(c,h_i,h,z)|^2\,,
\ee
where the``basis" functions ${\cal }F(c,h_i,h,z)$ (and ${\cal }F(c,\bar{h}_i,\bar{h},\bar{z})$) are known as Virasoro blocks, and encode the contributions to the four-point function from the whole Verma module corresponding to the primary with conformal weight $(h,\bar{h})$. They can be expanded in representations of the global conformal group, which would allow us to make direct contact with the analogous higher dimensional conformal block expansion, but in this paper we are more interested in study the expansion in the ``elliptic basis", obtained from mapping the branched sphere into an elliptic curve. 
\subsection{Elliptic expansion} 
In this section we want to present the expansion of conformal blocks on a elliptic curve which is equivalent to a double cover of a branched sphere \footnote{For details, look at  the analysis in section 7 of \cite{Maldacena:2015iua}.}. 

From general considerations, the original blocks ${\cal }F(c,h_i,h,z)$ have branches at the positions of the primary fields, namely at $\{\infty,1,z,0\}$. The map 
\be
q=e^{\pi i \tau},\qquad \tau={i \frac{K(1-z)}{K(z)}},
\ee
provides a conformal transformation into the upper-half plane ${\rm Im}(\tau)>0$ in such a way that the conformal block is a single valued function there, and therefore a series expansion in $q$ converges uniformly in ${\rm Im}(\tau)>0$ \cite{Zamolodchikov:1985ie}.

The variable $\tau$ corresponds to the modulus of a torus given by a double-cover of the Riemann sphere branched at $\{\infty,1,z,0\}$. This torus in turns can be described as an elliptic curve with the same branches as,
\be
y^2=x(z-x)(1-x)\,.
\ee
We can compute CFT correlators over  the torus in the usual way, by taking expectations values with a partition function given by, 
\be 
{\rm Tr}\left(e^{\pi \tau (L_0-{c\over 24})}e^{-\pi\bar{\tau}(\bar{L}_0-{c\over 24})}\right)={\rm Tr}\left(q^{L_0-{c\over 24}}\bar{q}^{\bar{L}_0-{c\over 24}}\right)\,.
\ee

Finally, we can then write an expansion for  $G(z,\bar{z})$ on the elliptic curve as,
\be
 G(z,\bar{z})=\Lambda(z)\Lambda(\bz)g(q,\bar{q})\,,
\ee
where 
\be\label{weyl_anomaly}
 \Lambda(z)=z^{\frac{c-1}{24}-h_{1}-h_{2}}\left(1-z\right)^{\frac{c-1}{24}-h_{2}-h_{3}}[\theta_{3}\left(q\right)]^{\frac{c-1}{2}-4\sum_{i=1}^{4}h_{i}}\,,
\ee
and $\theta_3(q)$ is a Jacobi Theta function. $g(q,\bar{q})$ is now expanded in a basis of blocks on the torus as,
\be
g(q,\bar{q})=\sum_{h,\bar{h}}C_{12,h, \bar{h}}C^{h, \bar{h}}_{~34}{\cal V}_{h,h_i,c}(q){\cal V}_{\bar{h},\bar{h}_i,c}(\bar{q})\,,
\ee
with the Virasoro blocks over the elliptic curve given by,
\begin{equation}\label{virasoro0}
\mathcal{V}_{h,h_{i},c}(q)={\left(16q\right)^{h-\frac{c-1}{24}}}H\left(c,h_{i},h,q\right),
\end{equation} \label{eq:blockV}

This modified blocks can be computed by projecting the correlator into contributions to the trace from each Verma module, schematically,
\be
\mathcal{V}_{h,h_{j},c}(q)\sim\sum_{N}^{\infty}\sum_{\{\sum_i n_i l_i\}=N}{\langle {\cal O}_h|\left(\prod_{n_i,l_i}L_{n_i}^{l_i}\right)\prod_{j=1}^4{\cal O}_{h_j}(z_j)\left(\prod_{n_i,l_i}L_{-n_i}^{l_i}\right)|{\cal O}_h\rangle\over \langle {\cal O}_h|\left(\prod_{n_i,l_i}L_{n_i}^{l_i}\right)\left(\prod_{n_i,l_i}L_{-n_i}^{l_i}\right)|{\cal O}_h\rangle}\,q^N \,,
\ee
however this procedure is very cumbersome. A more convenient way to deal with the blocks is through Zamolodchikov's  recursion relations \cite{Zamolodchikov:1987,Zamolodchikov:1985ie} which we proceed to describe in the next section.

\section{Zamolodchikov's recursion relations}
In two beautiful papers \cite{Zamolodchikov:1987,Zamolodchikov:1985ie} Zamolodchikov realized that generic blocks are strongly constrained by the existence of degenerate representations of the Virasoro algebra.  In particular, the function $H\left(c,h_{i},h,q\right)$ should have poles as a function of $h$ at points $h_{m,n}$ (defined below) corresponding to degenerate representations,
\be 
\mathcal{V}_{h,h_{i},c}(q) =  \sum_{m,n} \frac{ S_{m,n}}{h - h_{m,n}},
\ee
The residue $S_{m,n}$ of the pole at $h_{m,n}$ will be proportional to the block $
\mathcal{V}_{h_{mn}+mn,h_{i},c}(q)$ whose intermediate operator is evaluated on the degerate representation with dimension $h_{m,n}+mn$. These residues will have higher powers  $q^{mn}$ producing a series expansion in $q$.  All in all, this leads to the following recursion relation for $H$,
\be \label{recursionH}
H(b,h_i, h, q) = 1 + \sum_{m,n\ge1} \frac{q^{mn} R_{m,n}}{h - h_{m,n}} H(b,h_i, h_{m,n} + mn, q),
\ee
borrowing  the parametrization from Liouville theory, the central charge $c$, external operator dimensions $h_i$ and the degenerate operator dimensions $h_{mn}$  are written as,
\begin{equation}
Q=\left(b+\frac{1}{b}\right),\quad	c=1+6 Q^2,\quad  h_{i}=\frac{Q^2}{4}-\lambda_i^2,\quad h_{m,n}=\frac{Q^2}{4}-\lambda_{m,n}^2,
	\end{equation}
with 
\begin{equation}
\lambda_{m,n}=\frac{1}{2}\left(\frac{m}{b}+nb\right),
\end{equation}
and adopting the notation from \cite{Chen:2017yze}, $R_{m,n}$ is given by 
\begin{equation}\label{eq:Rmn}
R_{m,n}=2\frac{\prod_{p,q}\left(\lambda_1+\lambda_2-\lambda_{p,q}\right)\left(\lambda_1-\lambda_2-\lambda_{p,q}\right)\left(\lambda_3+\lambda_4-\lambda_{p,q}\right)\left(\lambda_3-\lambda_4-\lambda_{p,q}\right)}{\prod_{k,l}'\lambda_{k,l}},
\end{equation}
and the ranges of $p,q,k,$ and $l$ are:
\begin{align*}
p&=-m+1,-m+3,\cdots,m-3,m-1,\\
q&=-n+1,-n+3,\cdots,n-3,n-1,\\
k&=-m+1,-m+2,\cdots,m,\\
l&=-n+1,-n+2,\cdots,n. 
\end{align*}
with the prime in the denominator implying 
that $\left(k,l\right)=\{(0,0),(m,n)\}$ are excluded from the product.
The factor $R_{m,n}$ calls for a bit of faith and although it has passed several checks, mostly numerical (see for example \cite{Zamolodchikov:1995aa}), we are unaware of a definite prove of it yet. It is constructed in such way that it has zeros when an operator in the correlator belongs to the OPE between a degenerate state and any of the other external operators. 

Similarly to the above, Zamolodchikov derived an equivalent recursion relation by expanding the Virasoro blocks as sum over poles in the central charge $c$ instead of intermediate state dimension $h$ \cite{Zamolodchikov:1985ie}.

\subsection{Method of computing }

In order to simplify the analysis we will consider the simplest case with all external operator dimension equal to each other, i.e. $h_i = h_1$ for $i= 1\cdots 4$. At this configuration   $R_{m,n} = 0$ whenever $mn$ is odd. This means that every $H_{m,n}$ with odd $mn$ is also zero, as every term contributing to it contains at least one $R_{m_l,n_l}$ with odd $m_ln_l$. Therefore only even powers of $q$ will appear and we can forget about odd's $mn$. 

Let us start by noticing that by evaluating the left hand side of \eqref{recursionH} on the degenerate values $h=h_{mn}+mn$ we get a recursion for the coefficients,
\be \label{recursionHmn}
H(b,h_1, h_{mn}+mn, q) = 1 + \sum_{r,s\ge1} \frac{q^{rs} R_{r,s}}{h_{mn}+mn - h_{r,s}} H(b,h_1, h_{r,s} + rs, q)\,.
\ee
Now let us expand the function $H(b,h_1, h, q)$ and its coefficients \footnote{$H(b,h_1, h_{mn}, q)$ are coefficients as functions of $h$.} in a series expansion in $q$,
\be 
H(b,h_1, h, q)=\sum_{k\in 2\mathbb{N}} H^{(k)}(b,h_1, h)q^{k},\quad H(b,h_1, h_{mn}, q)=\sum_{{k\in 2\mathbb{N}}} H^{(k)}_{mn}(b,h_1,h_{mn})q^{k}\,.
\ee

Plugging this expansion in \eqref{recursionH} and \eqref{recursionHmn} respectively we find
\be\label{recursionH2}
	H^{(k)}(b,h_1, h) = \sum_{i\in 2\mathbb{N}}^{k\in 2\mathbb{N}} \sum_{\{m_ln_l=i\}} \frac{R_{m_l,n_l}}{h - h_{m_l,n_l}} H_{m_l,n_l}^{(k-i)}(b,h_1, h_{m_l,n_l}) ,
\ee
and
\be\label{recursionHmn2}
H^{(k)}_{m,n}(b,h_1, h_{mn}) = \sum_{i\in 2\mathbb{N}}^{k\in 2\mathbb{N}} \sum_{\{m_ln_l=i\}}\frac{ R_{m_l,n_l}}{h_{m,n}+mn-h_{m_l,n_l}}H_{m_l,n_l}^{(k-i)}(b,h_1, h_{m_l,n_l}) ,
\ee
where the notation $\{m_ln_l=i\}$ means the set of all pair of integers $\{m_ln_l\}$ whose product equals $i$.

From the two recursions above we see the strategy, for a given $k$, we expand \eqref{recursionH2} in terms of the coefficients $H_{m_l,n_l}^{(k-i)}$ and then we use \eqref{recursionHmn2} to write them in terms of lower order coefficients $H_{m_l,n_l}^{(k'<k-i)}$. Unfortunately, this becomes tedious very quickly, but is friendly enough to allow being implemented in a computer algebra algorithm (see \cite{Chen:2017yze} for a nice implementation), although it give not very illuminating expression at moderate higher orders.

\section{Large-h analysis}
Notice that due to the fact 
\be 
\lim_{h\to\infty}H(b,h_i, h, q)=1\,,
\ee
the prefactor in front of $H(b,h_i, h, q)$ in \eqref{virasoro0} corresponds to the $h\rightarrow \infty$ limit of $\CV_h$. Based on this observation, is natural to believe that the recursion relation might simplify in a series expansion around ${1\over h}$. Is the goal of this section to provide evidence that at leading order in ${1\over h}$, the recursion relation can be solved. 

Lets go back to the expansion \eqref{recursionH},
\be 
H(b,h_1, h, q) = 1 + \sum_{m,n\ge1} \frac{q^{mn} R_{m,n}}{h - h_{m,n}} H(b,h_1, h_{m,n} + mn, q),
\ee
the key observation here is that the residues, and in particular the coefficients $H(b,h_1, h_{m,n} + mn, q)$,  does not depend on $h$, henceforth assuming  that $h>>h_{m,n}$ for any $(m,n)$,\footnote{Notice that we are taking $c$ to be finite and then $h_{mn}$ will be small respect to $h$ as long as $(m,n)<< {\cal O}(h)$.} such as we can approximate the expansion above by,
\be \label{recursionHLargeh}
H(b,h_1, h, q) = 1 +\frac{1}{h}  \sum_{m,n\ge1} q^{mn} R_{m,n}\left(1+\frac{h_{mn}}{h}+\left(\frac{h_{mn}}{h}\right)^2+\cdots\right)H(b,h_1, h_{m,n} + mn, q),
\ee
we are interested into the leading term only, in other words, the expansion we want to deal with has the simplified form,
\be \label{recursionHLargehleading}
H(b,h_1, h, q) = 1 +\frac{1}{h}  \sum_{m,n\ge1} q^{mn} R_{m,n}H(b,h_1, h_{m,n} + mn, q)+ {\cal O}\left({1\over h^2}\right),
\ee
\subsection{Leading order}
Unfortunately the recursion for the coefficients \eqref{recursionHmn2} does not get affected by the limit and we still need to deal with the cumbersome long expression coming from several iterations. However, after evaluating some few terms with some involved algebra, the first few orders of $H(b,h_1, h, q)$ simplify more than we could have wished for. For a given order at the $q-$expansion \eqref{recursionHLargehleading} the corresponding contribution from all the contributing coefficients  become proportional to a single global function defined as,
\be\label{global_factor}
H_1(h_1,c)=\frac{1}{16}((c+1)-32 h_1)((c+5)-32 h_1)\,. 
\ee
By using an accordingly modified version of the algorithm developed in  \cite{Chen:2017yze} we were able to compute a few  low order terms up to $q^{18}$,
\beq\label{solution0}
H(b,h_1, h, q) &=& 1 +\frac{H_1(h_1,c)}{ h} \left(q^{2}+3q^{4}+4q^{6}+7 q^{8}+6q^{10}+12 q^{12}+8q^{14}+15q^{16}+13q^{18}\cdots \right)\nn\\
&&\quad+{\cal O}\left({1\over h^2}\right)\,.
\eeq
The sequence accompanying the $q$ expansion can be quickly recognized as generated by the sigma divisor function of order one, $\sigma_1(k)$, which gives the sum of all divisors of an integer $k$ \footnote{see table 24.6 \cite{abramowitz+stegun}}. 
Assuming that the pattern holds at higher orders, we can write the solution to the Virasoro block at leading order in a large$-h$ expansion as,
\be\label{solution}
H(b,h_1, h, q) = 1 +\frac{H_1(h_1,c)}{ h}  \sum_{k=1}^{\infty}q^{2k} \sigma_1(k)+{\cal O}\left({1\over h^2}\right)\,.
\ee
As a sanity check, we can immediately recognize that the only two solutions to the condition $H_1(h_1,c)=0$ are $\{c=1, h_1=1/16\}$ and $\{c=25, h_1=15/16\}$ which agrees with the results from \cite{Zamolodchikov:426555}, and the Virasoro blocks in that case are given by 
\be 
F\left(1,{1\over 16},h,z\right)={(16q)^{h}}(z(1-z))^{-{1\over 8}}\theta_3(q)^{-1}\,,
\ee
with a similar expression for the second solution.

The solution \eqref{solution} can be checked numerically to a reasonably high order by giving numerical values to $h_1$ and $c$  in order to compute a numerical solution to $H(b,h_1,h,q)$ as a function of $h$ from the numerical code \cite{Chen:2017yze}. Then expanding in $1/h$ and dividing each coefficient at a given order in the $q-$expansion by the numerical value of the global function \eqref{global_factor}, we can check that the left-over coefficients are indeed generated by the sigma divisor function up to order\footnote{We thank 
Shouvik Datta for performing this numerical check} $q^{500}$.

We can write the solution \eqref{solution} perhaps in a more illuminating form in terms of an Eisenstein series defined as (see for example \cite{Zagier08}),
\be\label{g2s}
G_{2s}(\tau)={1\over 2}\sum_{{m,n}\in \mathbb{Z}_2|(0,0)} \frac{1}{(m+n\tau)^{2s}}
\ee
which can be represented in terms of sigma divisors by the relation,
\be
G_{2s}(\tau)={(2\pi i)^{2s}\over(2 s -1)!}\left(-{B_{2s}\over 4 s}+\sum_{k=1}^{\infty} \sigma_{2s-1}(k)q^{2k}\right)\,,
\ee
here it is important to note that as a modular form the function $G_{2s}(\tau)$ only makes sense for $s> 1$, otherwise the $SL(2,\mathbb{Z})$-invariance would not hold, this is due to the fact that the series \eqref{g2s} is absolutely and uniformly convergent only for $s>1$, however for $s=1$ the sum can still be regulated. We discuss this issue in appendix. Nevertheless, an analogue of the holomorphic Eisenstein series can be defined even for $s = 1$, but it would give us a quasimodular form instead. By allowing  $s=1$ and using the definition $E_2(q)={6\over \pi^2}G_2(q)$, we can take \footnote{Another possibly useful representation of this solution might be in the form of a Lamber series 
\be
\sum_{k=1}^{\infty}q^{2k}  \sigma_1(k)=\sum_{k=1}^{\infty}\frac{k q^{2k}}{1-q^{2k}}\nn\,.
\ee},
\be
\sum_{k=1} \sigma_{1}(k)q^{2k}=\left(\frac{E_{2}(q)-1}{24}\right)\,,
\ee
and therefore rewrite the solution \eqref{solution} as,
\be\label{solution}
H(b,h_1, h, q) = 1 -\frac{H_1(h_1,c)}{ h} \left(\frac{E_{2}(q)-1}{24}\right)\,.
\ee
In this form, the leading order contribution seems to agree with what is expected from intuition. Due to the fact that we have written the Virasoro block on a elliptic curve, it has to be covariant under the fractional linear transformations preserving the lattice structure from the torus, i,e.
\be
 H\left({a z+b\over c z + d}\right)=(c z +d)^w H(z)\,,
\ee
this invariance in particular implies that the Virasoro blocks over the torus should be periodic $H(z+1)=H(z)$ and therefore have a Fourier expansion of the form
\be
H(z)=\sum_{k\in \mathbb{Z}}H^{(k)}q^{2k}\,,
\ee
with $q=e^{\pi i \tau}$ and
which is equivalently a consequence of the recursion relation itself.
Finally, automorphic forms on $SL(2,\mathbb{R})$ that are invariant under the action of the subgroup $SL(2,\mathbb{Z})$, can be written in terms of Eisenstein series. It is however puzzling for us, that the solution is given by a quasi-modular form instead of a modular form. An explanation of this emerge more clearly when looking through the light of supersymmetric gauge theories, as we will see in the next subsection.

\subsection{Comparison with instanton partition function of ${\cal N}=2$ SU(2) theory }
According to AGT correspondece \cite{Alday:2009aq}, Virasoro blocks corresponding to four-point correlation functions over the sphere are related to Nekrasov's instanton partitions functions \cite{Nekrasov:2002qd} for ${\cal N}=2$ SU(2) theory with ${\cal N}_f=4$. The dictionary mapping the gauge theory parameters into conformal field theory is given by,
\be  \label{dictionary}
b= \sqrt{\frac{\epsilon_2}{\epsilon_1}} \,, \quad
h_i = \frac{Q^2}{4} - \frac{M_i^2}{\epsilon_1 \epsilon_2} \,, 
\quad 
h = \frac{Q^2}{4} - \frac{a^2}{\epsilon_1 \epsilon_2} \,,
\ee
where
\be
M_1={m_3-m_2 \over 2}\,,\quad M_2={m_1-m_4 \over 2}\,,\quad M_3={m_3+m_2 \over 2}\,,\quad M_4={m_1+m_4 \over 2}\,,
\ee
$m_i$ corresponding to the masses of the hypermultiplets,  $\epsilon_i$ being deformation parameter need in the computation of Nekrasov's partition functions and $a$ the vector adjoint scalar vacuum expectation value \footnote{For details on the precise meaning of these parameters in the ${\cal N}=2$ SU(2) theory, please consult  \cite{Alday:2009aq}. }. Similarly as our analysis for Virasoro blocks, we want to express the instanton partition function $Z_{inst}$ as series in $h^{-1}$. It results more convenient to write the partition function expansion in terms of the ``free-energy" or pre-potential as is known in supersymmetric gauge theories, defined as,
\be
 F=-\epsilon_1\epsilon_2 \log Z_{inst}\,,
\ee
and explicitly AGT tell us,
\be
\Lambda(z)\mathcal{V}_{h,h_{i},c}(q)=Z_{inst}=e^{-\epsilon_1\epsilon_2 F}\,.
\ee
The desired expansion in inverse powers of $h$ for the Virasoro blocks should comes from a corresponding expansion for the free energy,
\be\label{Fexpansion}
F=-\pi i \tau h\,-R \log\theta_3(q)-\sum_{\ell=1}^{\infty}{1\over  2^{\ell+1}\, \ell}{f_{\ell}\over h^{\ell}}\,,
\ee
The first two terms can be rapidly computed from the CFT side by taking the logarithm at the global factor \eqref{weyl_anomaly} and the factor in front of $H(c,h_i,h,q)$ in \eqref {virasoro0} . The remaining constants $f_{\ell}$ have been considered in \cite{Billo:2013fi, Kashani-Poor:2013oza} up to order $\ell=3$. Here we quote only the first two orders, namely $f_1$  and $f_2$ from  \cite{Billo:2013fi}:\footnote{There is a typo in equation (3.20) and (3.21) of \cite{Billo:2013fi}, $T_1$ and $T_2$ should be swapped. }, 
\beq
f_1&=&{1\over 24} (4R-s^2+p)(4R-s^2+3p)E_2(q)-4(T_2\theta_4^4(q)-T_2\theta_2^4(q))\\
f_2&=&\frac{1}{144}\big(4R - s^2 +p\big)\big(4R - s^2 +3p\big)\big(4R - s^2 +4p\big)\,E_2^2
\notag\\&&~
-\frac{4}{3}\big(4R -s^2 +4p\big)\big(T_1 \,\theta_4^4-T_2 \,\theta_2^4\big)\,E_2\notag\\
&&~+\frac{1}{720}\Big[64 R^3-80 R^2(3s^2-4p)+4R(27s^4-88s^2p+49p^2)\nn\\
&&~~~~~~~~~~~~-13s^6+68s^4p-94s^2p^2+30p^3+2304 N\Big]\,E_4\nn\\
&&~-\frac{8}{3}\big(R -s^2 +p\big)\Big[T_1 \,\theta_4^4\big(2\theta_2^4+\theta_4^4\big)
+T_2 \,\theta_2^4\big(\theta_2^4+2\theta_4^4\big)\Big]~. 
\eeq
Here $R$, $T_\ell$ and $N$ are respectively, quadratic, quartic and sextic  polynomials in the masses $m_i$:
\begin{equation}
 \label{invariants}
  \begin{aligned}
     R&= \frac{1}{2}\,\sum_i m_i^2~,\\
     T_1&=\frac{1}{12}\,\sum_{i<j}m_i^2m_j^2-\frac{1}{24}\,\sum_im_i^4~,\\
     T_2&=-\frac{1}{24}\,\sum_{i<j}m_i^2m_j^2+\frac{1}{48}\,\sum_im_i^4-\frac{1}{2}\,m_1m_2m_3m_4~,\\ 
     N&=\frac{3}{16}\sum_{i<j<k}m_i^2m_j^2m_k^2-\frac{1}{96}
\,\sum_{i\not=j}m_i^4m_j^2+\frac{1}{96}\,\sum_i m_i^6 
 \end{aligned}
\end{equation}
By using  the dictionary \eqref{dictionary}, $f_1$  can be rewritten as,

\beq
f_1&=&\frac{1}{96} (c+1-8( h_1+ h_2+ h_3+ h_4) )(c+5-8( h_1+ h_2+ h_3+ h_4))E_2(q) \nn\\
&&+\frac{4}{3}
   \left({\over}\theta_2(q)^4 (h_1 (h_2-2 h_3+h_4)+h_2 (h_3-2 h_4)+h_3 h_4)\right.\nn\\
   &&\quad\left.-\theta_4(q)^4 (h_1 (-2 h_2+h_3+h_4)+h_2 (h_3+h_4)-2 h_3 h_4){\over}\right)\,,
\eeq
this expression (times a factor $\ell^{-1}\,2^{-\ell-1}$ coming from the normalization of the coeffficients in \eqref{Fexpansion}) generalizes the $q-$dependence of \eqref{solution}  to arbitrary external conformal dimensions. We have check numerically that the $q-$expansion of this formula, matches the $q-$expansion from Zamolodchikov's recursion relation up to order 20 \footnote{Here again we have used the code developed in \cite{Chen:2017yze}, which is specialized for the case $h_1=h_2$ and $h_3=h_4$,  and therefore the numerical check is restricted to that case. Also the checking order $q^{20}$ has been chosen arbitrarily and we can in principle perform this numerical check up to very high order only limited by machine computational  power.}, so in a way, we can thought of our computation in this paper from the Zamolodchikov's recursion relations, as yet another check for the AGT correspondence from the purely CFT perspective. In turn, the second order correction maps to,

\beq\label{second_order}
&&f_2=\nn\\
&&\frac{1}{720} \left[\frac{5E_2(q)^2}{8}  (c-8 (h_1+ h_2+ h_3+ h_4)+1) (c-8 (h_1+ h_2+ h_3+h_4)+5) (c-8( h_1+ h_2+ h_3+ h_4)+7)\right.\nn\\
  &&\left. -160 E_2(q) (c-8( h_1+ h_2+ h_3+ h_4)+7)
   \left(\theta_2^4 (-h_1 (h_2-2 h_3+h_4)-h_2 (h_3-2 h_4)-h_3 h_4)\right.\right.\nn\\
   &&\left.\left.+\theta_4^4 (h_1 (-2 h_2+h_3+h_4)+h_2 (h_3+h_4)-2 h_3 h_4)\right)\right.\nn\\
   &&\left.+E_4(q) \left(\frac{8}{27} (c-6 (h_1+ h_2+ h_3+
   h_4)-1)^3-\frac{10}{9} (c-9) (-c+6( h_1+ h_2+h_3+
   h_4)+1)^2\right.\right.\nn\\
   &&\left.\left.+\frac{1}{18} (c (9 c-194)+773) (c-6 (h_1+ h_2+
   h_3+ h_4)-1)-\frac{13 (c-1)^4}{1296}+\frac{17}{9} (c-1)^2\right.\right.\nn\\
   &&\left.\left.-\frac{47
   (c-1)}{3}+2304 N+30\right)\right.\nn\\
   &&\left.-1920 \theta_4(q)^4
   (h_1+h_2+h_3+h_4-1) \left(\theta_2(q)^4
   (h_1-h_3) (h_2-h_4)+\theta_4(q)^4
   (h_1-h_2) (h_3-h_4)\right)\right]
\eeq
where the parameter $N$ in the line before the last, corresponds to the sextic polynomial in masses, and translate to a big function in CFT parameters that is displayed in the appendix to made the formulas in this sections more compact.
\subsection{Some application}
Within the bootstrap framework in higher dimensions, it is well known that the light cone limit for the OPE expansion of the vacuum, is controlled by exchange operators at large spin \cite{Komargodski:2012ek, Fitzpatrick:2014vua}. This has lead to the development of a large spin perturbation theory \cite{Alday:2016njk, Alday:2015ewa} whose asymptotic series can be resummed  from a brilliant inversion formula \cite{Caron-Huot:2017vep, Simmons-Duffin:2017nub}. We believe the results from this paper can be use to perform equivalent analysis in two dimensional conformal field theories,  as those done for example in the computation of anomalous dimension at large spin in \cite{Cardona:2018qrt, Cardona:2018dov, Liu:2018jhs}. More explicitly, on the one hand the large spin limit can be taken by fixing, say $\bar{h}$, and taking $h$ to be large. On the other hand, by a saddle point analysis (see for example \cite{Kraus:2016nwo}) is straightforward to see that the large spin limit is controlled in the crossed channel by a small $\tau$, and therefore, by using the expansion of Virasoro blocks in $h^{-1}$ we can compute corrections to the crossed blocks as an expansion in small $\tau$.

This program has already been started very recently in \cite{Collier:2018exn, Collier:2019weq} by using  the Fusion kernel instead of conformal blocks, and therefore it would be interesting to understand those new results from the perspective of Virasoro blocks at large dimension and large spin.

\appendix
\section{The Eisenstein series of weight two}
The Eisenstein series of weight two, defined as in \eqref{g2s} for  $s=1$, transforms in the following way under $SL(2,\mathbb{Z})$ transformations,
\be
G_2\left({az+b\over cz+d}\right) =(c z+d)^2 G_2(z)-\pi i c(cz+d)\,.
\ee
Quasi-modularity is manifested in the additional ``anomalous" piece at the right hand side. The series  \eqref{g2s} for  $s=1$ does not converge absolutely, but just at the edge of convergence, in other words, it still converges at $s=1+\epsilon$,  for arbitrarily small $\epsilon>0$. Let us introduce,
\be
G_{2, \epsilon}(\tau)={1\over 2}\sum'_{{m,n}} \frac{1}{(m\tau+n)^{2}|m\tau+n|^{2\epsilon}}
\ee
where the prime on the summation symbol means $(m,n)=(0,0)$ is to be omitted. This series transforms as a modular form under $SL(2,\mathbb{Z})$, i.e,
\be
G_{2,\epsilon}\left({az+b\over cz+d}\right) =(c z+d)^2 |c z+d|^{2\epsilon}G_{2,\epsilon}\,,
\ee
it can be proved that the limit when $\epsilon$ goes to zero exists and is given by \footnote{For a prove of this statement and basics about (Quasi-) Modular forms, see for example \cite{Zagier08}},
\be
\lim_{\epsilon\to 0}G_{2, \epsilon}(\tau)=G_{2}(\tau)-{\pi\over 2 {\rm Im}(\tau)}\,.
\ee
This implies, that the  Eisenstein series of weight two can be regulated and therefore modularity can be preserved, at the expense of holomorphicity. 
\section{Sextic polynomial in masses}
Here we show explicitly the quantity $N$ appearing in equation \eqref{second_order},

\beq
\small
N&=&\frac{1}{32} \left((c-12 h_2-12 h_4-1) h_1^2-2 \left(6
   h_2^2-12 (h_3+h_4) h_2+6 h_4^2+h_3 (c-12
   h_4-1)\right) h_1\right.\nn\\
   &&\left.+c h_3^2-h_3^2+c h_4^2-12
   h_3 h_4^2-h_4^2+h_2^2 (c-12 h_3-1)-12 h_3^2
   h_4-2 h_2 \left(6 h_3^2-12 h_4 h_3+(c-1)
   h_4\right)\right)\nn\\
   &&+\frac{1}{96} \left(\frac{20736
   (h_1-h_3)^8}{\left(-c+12 h_1+12 h_3+\sqrt{-c+24
   h_1+1} \sqrt{-c+24 h_3+1}+1\right)^4}\right.\nn\\
   &&\left.+\frac{\left(-c+12 h_1+12
   h_3+\sqrt{-c+24 h_1+1} \sqrt{-c+24
   h_3+1}+1\right)^4}{20736}\right.\nn\\
   &&\left.+\frac{\left(-c+12 h_2+12
   h_4+\sqrt{-c+24 h_2+1} \sqrt{-c+24
   h_4+1}+1\right)^4}{20736}\right.\nn\\
   &&\left.+\frac{20736 (h_2-h_4)^8}{\left(-c+12
   h_2+12 h_4+\sqrt{-c+24 h_2+1} \sqrt{-c+24
   h_4+1}+1\right)^4}\right)\nn\\
   &&+\frac{1}{768} \left(\frac{96 \left(-c+12
   h_2+12 h_4+\sqrt{-c+24 h_2+1} \sqrt{-c+24 h_4+1}+1\right)
   (h_1-h_3)^4}{\left(-c+12 h_1+12 h_3+\sqrt{-c+24
   h_1+1} \sqrt{-c+24 h_3+1}+1\right)^2}\right.\nn\\
   &&\left.+\frac{96
   (h_1-h_3)^4}{-c+12 h_1+12 h_3+\sqrt{-c+24 h_1+1}
   \sqrt{-c+24 h_3+1}+1}\right.\nn\\
   &&\left.-\frac{13824 (h_2-h_4)^2
   (h_1-h_3)^4}{\left(-c+12 h_1+12 h_3+\sqrt{-c+24
   h_1+1} \sqrt{-c+24 h_3+1}+1\right)^2 }\right.\nn\\
   &&\times\left.\frac{1}{\left(c-12 h_2-12
   h_4-\sqrt{-c+24 h_2+1} \sqrt{-c+24 h_4+1}-1\right)}\right.\nn\\
   &&\left.+\frac{2
   \left(-c+12 h_2+12 h_4+\sqrt{-c+24 h_2+1} \sqrt{-c+24
   h_4+1}+1\right)^2 (h_1-h_3)^2}{-3 c+36 h_1+36 h_3+3
   \sqrt{-c+24 h_1+1} \sqrt{-c+24 h_3+1}+3}\right.\nn\\
   &&\left.+\frac{2}{3} \left(-c+12
   h_1+12 h_3+\sqrt{-c+24 h_1+1} \sqrt{-c+24 h_3+1}+1\right)
   (h_1-h_3)^2\right.\nn\\
   &&\left.+\frac{13824 (h_2-h_4)^4
   (h_1-h_3)^2}{\left(-c+12 h_1+12 h_3+\sqrt{-c+24
   h_1+1} \sqrt{-c+24 h_3+1}+1\right)}\right.\nn\\
   &&\times\left.\frac{1}{ \left(-c+12 h_2+12
   h_4+\sqrt{-c+24 h_2+1} \sqrt{-c+24
   h_4+1}+1\right)^2}\right.\nn\\
   &&\left.+\frac{1}{216} \left(-c+12 h_1+12
   h_3+\sqrt{-c+24 h_1+1} \sqrt{-c+24 h_3+1}+1\right) \right.\nn\\
   &&\times\left.\left(-c+12
   h_2+12 h_4+\sqrt{-c+24 h_2+1} \sqrt{-c+24
   h_4+1}+1\right)^2\right.\nn\\
   &&\left.-\frac{1}{216} \left(-c+12 h_1+12
   h_3+\sqrt{-c+24 h_1+1} \sqrt{-c+24 h_3+1}+1\right)^2 \right.\nn\\
   &&\times\left.\left(c-12
   h_2-12 h_4-\sqrt{-c+24 h_2+1} \sqrt{-c+24
   h_4+1}-1\right)\right.\nn\\
   &&\left.+\frac{2}{3} (h_2-h_4)^2 \left(-c+12 h_2+12
   h_4+\sqrt{-c+24 h_2+1} \sqrt{-c+24 h_4+1}+1\right)\right.\nn\\
   &&\left.+\frac{96
   (h_2-h_4)^4}{-c+12 h_2+12 h_4+\sqrt{-c+24 h_2+1}
   \sqrt{-c+24 h_4+1}+1}\right.\nn\\
   &&\left.+\frac{2 \left(-c+12 h_1+12 h_3+\sqrt{-c+24
   h_1+1} \sqrt{-c+24 h_3+1}+1\right)^2 (h_2-h_4)^2}{-3 c+36
   h_2+36 h_4+3 \sqrt{-c+24 h_2+1} \sqrt{-c+24
   h_4+1}+3}\right.\nn\\
   &&\left.+\frac{96 \left(-c+12 h_1+12 h_3+\sqrt{-c+24
   h_1+1} \sqrt{-c+24 h_3+1}+1\right)
   (h_2-h_4)^4}{\left(-c+12 h_2+12 h_4+\sqrt{-c+24
   h_2+1} \sqrt{-c+24 h_4+1}+1\right)^2}\right)
\eeq
\section*{Acknowledgments}
The author thank Cindy Keeler, Shouvik Datta, Sridip Pal, Diptarka Das and Jan Troost for helpful discussions. I also would like to thank the anonymous referees for their insightful comments and suggestions. This work is supported by the U.S. Department of Energy under grant number DE-SC0019470.

\hspace{2cm}
\bibliographystyle{ieeetr}
\bibliography{VirasoroBib}

\begin{thebibliography}{10}

\bibitem{Belavin:1984vu}
A.~Belavin, A.~M. Polyakov, and A.~Zamolodchikov, ``{Infinite Conformal
  Symmetry in Two-Dimensional Quantum Field Theory},'' {\em Nucl. Phys. B},
  vol.~241, pp.~333--380, 1984.

\bibitem{Alday:2009aq}
L.~F. Alday, D.~Gaiotto, and Y.~Tachikawa, ``{Liouville Correlation Functions
  from Four-dimensional Gauge Theories},'' {\em Lett. Math. Phys.}, vol.~91,
  pp.~167--197, 2010.

\bibitem{Fateev:2009aw}
V.~Fateev and A.~Litvinov, ``{On AGT conjecture},'' {\em JHEP}, vol.~02,
  p.~014, 2010.

\bibitem{Barnich:2009se}
G.~Barnich and C.~Troessaert, ``{Symmetries of asymptotically flat 4
  dimensional spacetimes at null infinity revisited},'' {\em Phys. Rev. Lett.},
  vol.~105, p.~111103, 2010.

\bibitem{Kapec:2014opa}
D.~Kapec, V.~Lysov, S.~Pasterski, and A.~Strominger, ``{Semiclassical Virasoro
  symmetry of the quantum gravity $ \mathcal{S}$-matrix},'' {\em JHEP},
  vol.~08, p.~058, 2014.

\bibitem{Cardona:2017keg}
C.~Cardona and Y.-t. Huang, ``{S-matrix singularities and CFT correlation
  functions},'' {\em JHEP}, vol.~08, p.~133, 2017.

\bibitem{Ponsot:1999uf}
B.~Ponsot and J.~Teschner, ``{Liouville bootstrap via harmonic analysis on a
  noncompact quantum group},'' 11 1999.

\bibitem{Ponsot:2000mt}
B.~Ponsot and J.~Teschner, ``{Clebsch-Gordan and Racah-Wigner coefficients for
  a continuous series of representations of U(q)(sl(2,R))},'' {\em Commun.
  Math. Phys.}, vol.~224, pp.~613--655, 2001.

\bibitem{Zamolodchikov:426555}
A.~B. Zamolodchikov, ``{Two-dimensional conformal symmetry and critical
  four-spin correlation functions in the Ashkin-Teller model},'' {\em Sov.
  Phys. - JETP}, vol.~63, pp.~1061--1066, 1986.

\bibitem{Runkel:2001ng}
I.~Runkel and G.~Watts, ``{A Nonrational CFT with c = 1 as a limit of minimal
  models},'' {\em JHEP}, vol.~09, p.~006, 2001.

\bibitem{Gamayun:2012ma}
O.~Gamayun, N.~Iorgov, and O.~Lisovyy, ``{Conformal field theory of Painlev\'e
  VI},'' {\em JHEP}, vol.~10, p.~038, 2012.
\newblock [Erratum: JHEP 10, 183 (2012)].

\bibitem{Iorgov:2013uoa}
N.~Iorgov, O.~Lisovyy, and Y.~Tykhyy, ``{Painlevé VI connection problem and
  monodromy of $c=1$ conformal blocks},'' {\em JHEP}, vol.~12, p.~029, 2013.

\bibitem{Iorgov:2014vla}
N.~Iorgov, O.~Lisovyy, and J.~Teschner, ``{Isomonodromic tau-functions from
  Liouville conformal blocks},'' {\em Commun. Math. Phys.}, vol.~336, no.~2,
  pp.~671--694, 2015.

\bibitem{Litvinov:2013sxa}
A.~Litvinov, S.~Lukyanov, N.~Nekrasov, and A.~Zamolodchikov, ``{Classical
  Conformal Blocks and Painleve VI},'' {\em JHEP}, vol.~07, p.~144, 2014.

\bibitem{Alba:2010qc}
V.~A. Alba, V.~A. Fateev, A.~V. Litvinov, and G.~M. Tarnopolskiy, ``{On
  combinatorial expansion of the conformal blocks arising from AGT
  conjecture},'' {\em Lett. Math. Phys.}, vol.~98, pp.~33--64, 2011.

\bibitem{Perlmutter:2015iya}
E.~Perlmutter, ``{Virasoro conformal blocks in closed form},'' {\em JHEP},
  vol.~08, p.~088, 2015.

\bibitem{Fitzpatrick:2015foa}
A.~L. Fitzpatrick, J.~Kaplan, M.~T. Walters, and J.~Wang, ``{Hawking from
  Catalan},'' {\em JHEP}, vol.~05, p.~069, 2016.

\bibitem{Fitzpatrick:2015qma}
A.~L. Fitzpatrick, J.~Kaplan, M.~T. Walters, and J.~Wang, ``{Eikonalization of
  Conformal Blocks},'' {\em JHEP}, vol.~09, p.~019, 2015.

\bibitem{Fitzpatrick:2015zha}
A.~L. Fitzpatrick, J.~Kaplan, and M.~T. Walters, ``{Virasoro Conformal Blocks
  and Thermality from Classical Background Fields},'' {\em JHEP}, vol.~11,
  p.~200, 2015.

\bibitem{Fitzpatrick:2014vua}
A.~Fitzpatrick, J.~Kaplan, and M.~T. Walters, ``{Universality of Long-Distance
  AdS Physics from the CFT Bootstrap},'' {\em JHEP}, vol.~08, p.~145, 2014.

\bibitem{Fitzpatrick:2016mtp}
A.~L. Fitzpatrick, J.~Kaplan, D.~Li, and J.~Wang, ``{Exact Virasoro Blocks from
  Wilson Lines and Background-Independent Operators},'' {\em JHEP}, vol.~07,
  p.~092, 2017.

\bibitem{Hijano:2015qja}
E.~Hijano, P.~Kraus, E.~Perlmutter, and R.~Snively, ``{Semiclassical Virasoro
  blocks from AdS$_{3}$ gravity},'' {\em JHEP}, vol.~12, p.~077, 2015.

\bibitem{Fitzpatrick:2015dlt}
A.~L. Fitzpatrick and J.~Kaplan, ``{Conformal Blocks Beyond the Semi-Classical
  Limit},'' {\em JHEP}, vol.~05, p.~075, 2016.

\bibitem{Beccaria:2015shq}
M.~Beccaria, A.~Fachechi, and G.~Macorini, ``{Virasoro vacuum block at
  next-to-leading order in the heavy-light limit},'' {\em JHEP}, vol.~02,
  p.~072, 2016.

\bibitem{Besken:2019jyw}
M.~Becsken, S.~Datta, and P.~Kraus, ``{Semi-classical Virasoro blocks: proof of
  exponentiation},'' {\em JHEP}, vol.~01, p.~109, 2020.

\bibitem{Ginsparg:1988ui}
P.~H. Ginsparg, ``{APPLIED CONFORMAL FIELD THEORY},'' in {\em {Les Houches
  Summer School in Theoretical Physics: Fields, Strings, Critical Phenomena}},
  pp.~1--168, 9 1988.

\bibitem{di1996conformal}
P.~Di~Francesco, P.~Mathieu, and D.~S{\'e}n{\'e}chal, {\em Conformal Field
  Theory}.
\newblock Graduate texts in contemporary physics, Island Press, 1996.

\bibitem{Zamolodchikov:1987}
A.~Zamolodchikov, ``{Conformal symmetry in two-dimensional space: Recursion
  representation of conformal block},'' {\em Theoret. and Math. Phys.},
  vol.~73, pp.~1088--1093, 1987.

\bibitem{Zamolodchikov:1985ie}
A.~Zamolodchikov, ``{CONFORMAL SYMMETRY IN TWO-DIMENSIONS: AN EXPLICIT
  RECURRENCE FORMULA FOR THE CONFORMAL PARTIAL WAVE AMPLITUDE},'' {\em Commun.
  Math. Phys.}, vol.~96, pp.~419--422, 1984.

\bibitem{Hadasz:2009db}
L.~Hadasz, Z.~Jaskolski, and P.~Suchanek, ``{Recursive representation of the
  torus 1-point conformal block},'' {\em JHEP}, vol.~01, p.~063, 2010.

\bibitem{Poghossian:2009mk}
R.~Poghossian, ``{Recursion relations in CFT and N=2 SYM theory},'' {\em JHEP},
  vol.~12, p.~038, 2009.

\bibitem{Fateev:2009me}
V.~A. Fateev, A.~Litvinov, A.~Neveu, and E.~Onofri, ``{Differential equation
  for four-point correlation function in Liouville field theory and elliptic
  four-point conformal blocks},'' {\em J. Phys. A}, vol.~42, p.~304011, 2009.

\bibitem{Alkalaev:2016fok}
K.~Alkalaev, R.~Geiko, and V.~Rappoport, ``{Various semiclassical limits of
  torus conformal blocks},'' {\em JHEP}, vol.~04, p.~070, 2017.

\bibitem{Chen:2017yze}
H.~Chen, C.~Hussong, J.~Kaplan, and D.~Li, ``{A Numerical Approach to Virasoro
  Blocks and the Information Paradox},'' {\em JHEP}, vol.~09, p.~102, 2017.

\bibitem{Kashani-Poor:2013oza}
A.-K. Kashani-Poor and J.~Troost, ``{Transformations of Spherical Blocks},''
  {\em JHEP}, vol.~10, p.~009, 2013.

\bibitem{Billo:2013fi}
M.~Billo, M.~Frau, L.~Gallot, A.~Lerda, and I.~Pesando, ``{Deformed N=2
  theories, generalized recursion relations and S-duality},'' {\em JHEP},
  vol.~04, p.~039, 2013.

\bibitem{KashaniPoor:2012wb}
A.-K. Kashani-Poor and J.~Troost, ``{The toroidal block and the genus
  expansion},'' {\em JHEP}, vol.~03, p.~133, 2013.

\bibitem{Kashani-Poor:2014mua}
A.-K. Kashani-Poor and J.~Troost, ``{Quantum geometry from the toroidal
  block},'' {\em JHEP}, vol.~08, p.~117, 2014.

\bibitem{Huang:2011qx}
M.-x. Huang, A.-K. Kashani-Poor, and A.~Klemm, ``{The $\Omega$ deformed B-model
  for rigid $\mathcal{N}=2$ theories},'' {\em Annales Henri Poincare}, vol.~14,
  pp.~425--497, 2013.

\bibitem{Billo:2011pr}
M.~Billo, M.~Frau, L.~Gallot, and A.~Lerda, ``{The exact 8d chiral ring from 4d
  recursion relations},'' {\em JHEP}, vol.~11, p.~077, 2011.

\bibitem{Collier:2019weq}
S.~Collier, A.~Maloney, H.~Maxfield, and I.~Tsiares, ``{Universal Dynamics of
  Heavy Operators in CFT$_2$},'' 11 2019.

\bibitem{Cardona:2020-2}
C.~Cardona and C.~Keeler, ``{Universal asymptotics of conformal field theory
  from Liouville},'' 2020.

\bibitem{Brehm:2019pcx}
E.~M. Brehm and D.~Das, ``{Aspects of the S transformation Bootstrap},'' {\em
  J. Stat. Mech.}, vol.~2005, p.~053103, 2020.

\bibitem{Das:2017cnv}
D.~Das, S.~Datta, and S.~Pal, ``{Universal asymptotics of three-point
  coefficients from elliptic representation of Virasoro blocks},'' {\em Phys.
  Rev. D}, vol.~98, no.~10, p.~101901, 2018.

\bibitem{Jackson:2014nla}
S.~Jackson, L.~McGough, and H.~Verlinde, ``{Conformal Bootstrap, Universality
  and Gravitational Scattering},'' {\em Nucl. Phys. B}, vol.~901, pp.~382--429,
  2015.

\bibitem{Maldacena:2015iua}
J.~Maldacena, D.~Simmons-Duffin, and A.~Zhiboedov, ``{Looking for a bulk
  point},'' {\em JHEP}, vol.~01, p.~013, 2017.

\bibitem{Zamolodchikov:1995aa}
A.~B. Zamolodchikov and A.~B. Zamolodchikov, ``{Structure constants and
  conformal bootstrap in Liouville field theory},'' {\em Nucl. Phys. B},
  vol.~477, pp.~577--605, 1996.

\bibitem{abramowitz+stegun}
M.~Abramowitz and I.~A. Stegun, {\em Handbook of Mathematical Functions with
  Formulas, Graphs, and Mathematical Tables}.
\newblock New York: Dover, ninth dover printing, tenth gpo printing~ed., 1964.

\bibitem{Zagier08}
K.~Ranestad, {\em The 1-2-3 of Modular Forms}.
\newblock Springer Berlin Heidelberg, 2008.

\bibitem{Nekrasov:2002qd}
N.~A. Nekrasov, ``{Seiberg-Witten prepotential from instanton counting},'' {\em
  Adv. Theor. Math. Phys.}, vol.~7, no.~5, pp.~831--864, 2003.

\bibitem{Komargodski:2012ek}
Z.~Komargodski and A.~Zhiboedov, ``{Convexity and Liberation at Large Spin},''
  {\em JHEP}, vol.~11, p.~140, 2013.

\bibitem{Alday:2016njk}
L.~F. Alday, ``{Large Spin Perturbation Theory for Conformal Field Theories},''
  {\em Phys. Rev. Lett.}, vol.~119, no.~11, p.~111601, 2017.

\bibitem{Alday:2015ewa}
L.~F. Alday and A.~Zhiboedov, ``{An Algebraic Approach to the Analytic
  Bootstrap},'' {\em JHEP}, vol.~04, p.~157, 2017.

\bibitem{Caron-Huot:2017vep}
S.~Caron-Huot, ``{Analyticity in Spin in Conformal Theories},'' {\em JHEP},
  vol.~09, p.~078, 2017.

\bibitem{Simmons-Duffin:2017nub}
D.~Simmons-Duffin, D.~Stanford, and E.~Witten, ``{A spacetime derivation of the
  Lorentzian OPE inversion formula},'' {\em JHEP}, vol.~07, p.~085, 2018.

\bibitem{Cardona:2018qrt}
C.~Cardona, S.~Guha, S.~K. Kanumilli, and K.~Sen, ``{Resummation at finite
  conformal spin},'' {\em JHEP}, vol.~01, p.~077, 2019.

\bibitem{Cardona:2018dov}
C.~Cardona and K.~Sen, ``{Anomalous dimensions at finite conformal spin from
  OPE inversion},'' {\em JHEP}, vol.~11, p.~052, 2018.

\bibitem{Liu:2018jhs}
J.~Liu, E.~Perlmutter, V.~Rosenhaus, and D.~Simmons-Duffin, ``{$d$-dimensional
  SYK, AdS Loops, and $6j$ Symbols},'' {\em JHEP}, vol.~03, p.~052, 2019.

\bibitem{Kraus:2016nwo}
P.~Kraus and A.~Maloney, ``{A cardy formula for three-point coefficients or how
  the black hole got its spots},'' {\em JHEP}, vol.~05, p.~160, 2017.

\bibitem{Collier:2018exn}
S.~Collier, Y.~Gobeil, H.~Maxfield, and E.~Perlmutter, ``{Quantum Regge
  Trajectories and the Virasoro Analytic Bootstrap},'' {\em JHEP}, vol.~05,
  p.~212, 2019.

\end{thebibliography}
\end{document}